\documentclass[10pt,letterpaper]{article}
\usepackage[top=0.85in,left=2.75in,footskip=0.75in]{geometry}
\usepackage{amsmath,amssymb}
\usepackage{changepage}
\usepackage{textcomp,marvosym}
\usepackage{cite}
\usepackage{comment}
\usepackage{nameref,hyperref}
\usepackage[nopatch=eqnum]{microtype}
\DisableLigatures[f]{encoding = *, family = * }
\usepackage[table]{xcolor}
\usepackage{array}
\newcolumntype{+}{!{\vrule width 2pt}}
\newlength\savedwidth

\raggedright
\usepackage[aboveskip=1pt,labelfont=bf,labelsep=period,justification=raggedright,singlelinecheck=off]{caption}

\makeatletter
\renewcommand{\@biblabel}[1]{\quad#1.}
\makeatother
\usepackage{lastpage,fancyhdr,graphicx}
\usepackage{epstopdf}
\fancyheadoffset[L]{2.25in}
\fancyfootoffset[L]{2.25in}

\begin{document}
\vspace*{0.2in}

\begin{flushleft}
{\Large
\textbf\newline{Identifying topologically associating domains using differential kernels}
}
\newline
\\
Luka Maisuradze\textsuperscript{1},
Megan C. King\textsuperscript{2},
Ivan V. Surovtsev\textsuperscript{2},
Simon G. J. Mochrie\textsuperscript{3},
Mark D. Shattuck\textsuperscript{4},
Corey S. O'Hern\textsuperscript{5,3,6*},

\bigskip
\textbf{1} Department of Molecular Biophysics and Biochemistry, Yale University, New Haven, Connecticut, United States of America
\\
\textbf{2} Department of Cell Biology, Yale School of Medicine, New Haven, Connecticut, United States of America
\\
\textbf{3} Department of Physics, Yale University, New Haven, Connecticut, United States of America
\\
\textbf{4} Benjamin Levich Institute and Physics Department, The City College of New York, New York, New York, United States of America.
\\
\textbf{5} Department of Mechanical Engineering and Materials Science, Yale University,
New Haven, Connecticut, United States of America
\\
\textbf{6} Graduate Program in Computational Biology and Bioinformatics, Yale University,
New Haven, Connecticut, United States of America
\\
\bigskip

*corey.ohern@yale.edu

\end{flushleft}
\section*{Abstract}
Chromatin is a polymer complex of DNA and proteins that regulates gene expression. The three-dimensional (3D) structure and organization of chromatin controls DNA transcription and replication. High-throughput chromatin conformation capture techniques generate Hi-C maps that can provide insight into the 3D structure of chromatin. Hi-C maps can be represented as a symmetric matrix ${\cal A}_{ij}$, where each element represents the average contact probability or number of contacts between chromatin loci $i$ and $j$. Previous studies have detected topologically associating domains (TADs), or self-interacting regions in ${\cal A}_{ij}$ within which the contact probability is greater than that outside the region. Many algorithms have been developed to identify TADs within Hi-C maps. However, most TAD identification algorithms are unable to identify nested or overlapping TADs and for a given Hi-C map there is significant variation in the location and number of TADs identified by different methods. We develop a novel method to identify TADs, KerTAD, using a kernel-based technique from computer vision and image processing that is able to accurately identify nested and overlapping TADs. We benchmark this method against state-of-the-art TAD identification methods on both synthetic and experimental data sets. We find that the new method consistently has higher true positive rates (TPR) and lower false discovery rates (FDR) than all tested methods for both synthetic and manually annotated experimental Hi-C maps. The TPR for KerTAD is also largely insensitive to increasing noise and sparsity, in contrast to the other methods. We also find that KerTAD is consistent in the number and size of TADs identified across replicate experimental Hi-C maps for several organisms. Thus, KerTAD will improve automated TAD identification and enable researchers to better correlate changes in TADs to biological phenomena, such as enhancer-promoter interactions and disease states. 

\section*{Author summary}
Chromatin, which encodes the genetic information for cells, must fold into the cell nucleus that is many times smaller in size. The folded 3D structure of chromatin in the nucleus enables gene expression and proper cell function. With the advent of advanced chromatin conformation capture techniques, we can identify topologically associating domains (TADs), which are regions of the genome that prefer to interact within themselves rather than with neighboring regions. Numerous methods have been developed to automatically detect TADs in Hi-C maps, however, they frequently disagree on the location and number of TADs. We develop a new algorithm, KerTAD, to identify TADs using techniques from image processing and computer vision. We find that our method is more accurate on both synthetic and manually-annotated experimental Hi-C maps than all tested methods. Our method also performs well in the presence of noise and sparsity, which are frequently encountered in experimental Hi-C maps. KerTAD will enable future studies to elucidate the role of TADs in gene regulation and disease formation.

\section*{Introduction}

Chromatin is a polymer complex of DNA and proteins that forms chromosomes. Chromatin must undergo a highly organized compaction process to fit into the $\mu$m-sized nucleus. During this compaction process, chromatin forms hierarchical structures, such as loops, A/B compartments, and territories, across a range of length scales~\cite{bib1}~\cite{bib2}~\cite{bib3}~\cite{bib4}. The spatial organization of chromatin is essential for many nuclear processes, such as DNA replication and transcription. For example, during transcription, enhancer and promoter DNA regions that are separated on the chromatin fiber must come into close proximity through the formation of loops to increase the transcription of target genes~\cite{bib1,bib5}. Disruptions in chromatin loop formation can alter gene expression by preventing enhancer-promoter interactions~\cite{bib6,bib7}. To better understand the structural organization of chromatin, chromosome conformation capture and proximity ligation derivative techniques (in particular Hi-C) have been developed to elucidate genome-wide spatial interactions and structures~\cite{bib8}~\cite{bib9}. Hi-C generates an interaction matrix, $\mathcal{A}_{ij}$, where each element represents the frequency with which two loci $i$ and $j$ on chromatin are close in space, averaged over a cell population ~\cite{bib8}. Hi-C maps reveal significant interactions off the diagonal that are not expected for an extended polymer. In particular, Hi-C maps display topologically associating domains (TADs), or regions of increased self-interaction (with decreased interactions outside the region), typically presenting as a square of higher frequency centered on the diagonal~\cite{bib10,bib11}. TADs often indicate the formation, elongation, and dissolution of loops. Loops enable enhancer-promoter interactions and TAD boundaries are frequently enriched for insulator proteins and transcription marks, which explains why enhancer-promoter interactions occur mostly within TADs ~\cite{bib10,bib12,bib13,bib14,bib15,bib16}.

Several features of experimentally determined Hi-C maps, such as noise, sparsity, and low resolution, make TAD identification difficult. Further, TAD features are heterogeneous, e.g. while some TADs possess strong corner points and weak intensity in the interior of the TAD, others possess uniform intensity in the interior with weak borders. TADs are also often difficult to differentiate from the background power-law decay in the interaction frequency away from the diagonal that arises from expected distance-dependent polymer interactions~\cite{bib17}. The convention for TAD identification, or TAD calling, is to specify the starting and ending loci of each TAD in the interaction matrix $\mathcal{A}_{ij}$. However, TADs do not directly report on static chromatin structure, instead they provide a statistical description of dynamic chromatin organization that is influenced by the experimental methods used to construct the Hi-C maps\cite{bib12,bib18,bib51}. Currently, there is no ground-truth definition for TADs in Hi-C maps, and TAD definitions are scale- and resolution-dependent~\cite{bib12,bib18,bib19}. To illustrate this point, in Fig. 1A and 1B, we show the same segment (from $9$ to $13$ Mb) of mouse chromosome $17$ Hi-C map using both linear and logarithmic (base e) intensity scales, respectively. On the linear scale, TADs are not visible, whereas on the logarithmic scale, numerous overlapping and nested TADs appear. In Fig. 1C we show the same segment of mouse chromosome $17$ on a logarithmic scale, but from a different biological replicate, showing a much sparser Hi-C map and replicate to replicate fluctuations. 

\begin{figure}[!h] 
\includegraphics[width=\textwidth]{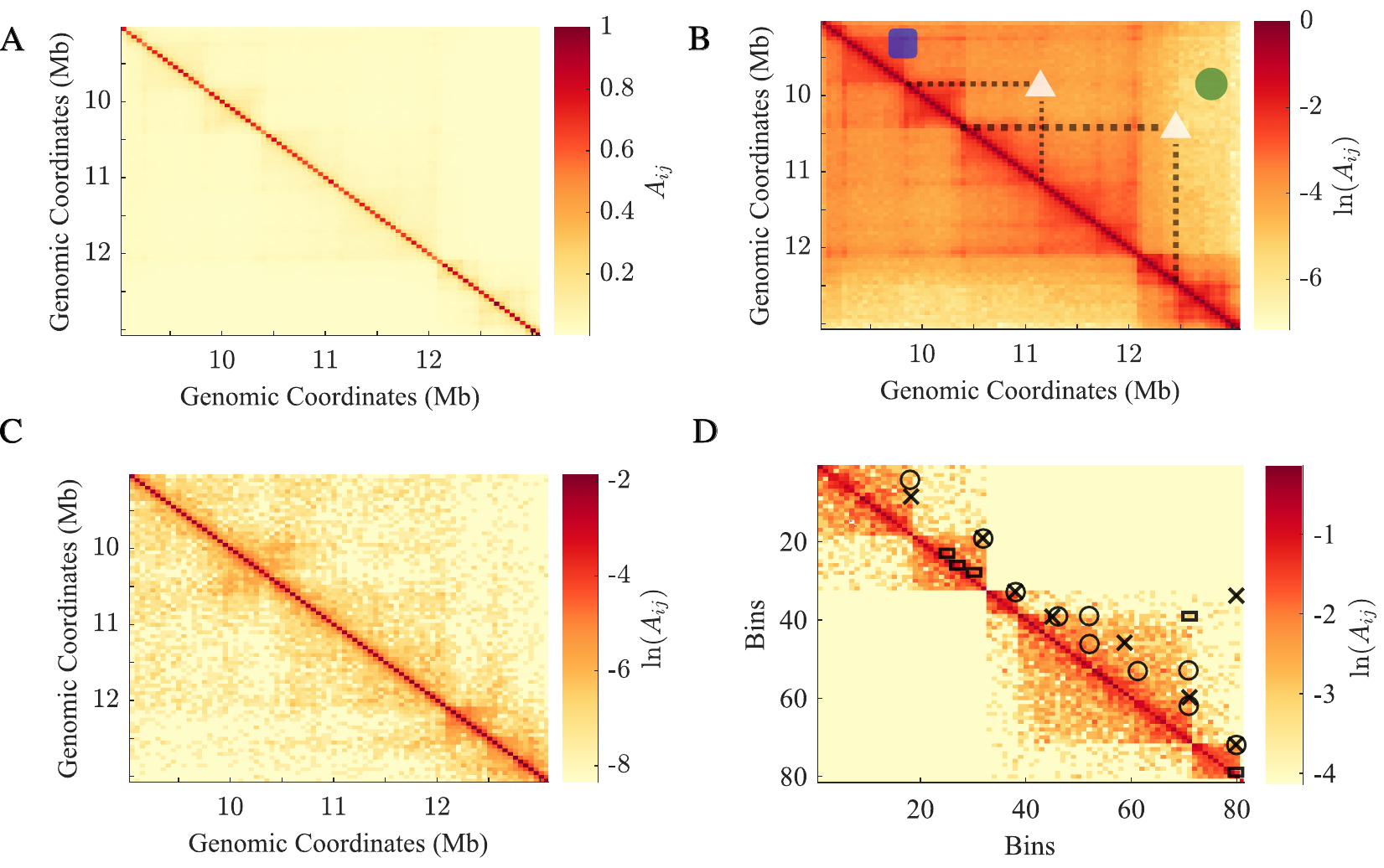}
\caption{{\bf Challenges of TAD identification on Hi-C maps.}
A: The choice of scale and normalization of Hi-C maps impacts the visibility of TADs. Mouse Hi-C map of chromosome $17$ (from $9$ to $13$ Mb) without preprocessing on a linear scale and normalized so that $0 \leq {\cal A}_{ij} \leq 1$ yields faint TADs. B: We show the same Hi-C map as in A, but plotted on a natural logarithmic scale. The blue square indicates the corner of a clear TAD and the dotted lines in the upper triangular matrix denote its boundaries. The green circle shows a region of noise near a TAD boundary, and the white triangles (and associated dashed lines) indicate borderline TADs that were not visible in the left image.  C: The same region of mouse chromosome $17$, but from a different biological replicate with many more low intensity values off the diagonal. D: A synthetic Hi-C map generated from negative binomial distribution sampling with TADs identified (shown in the upper right triangle) using three state-of-the-art TAD calling algorithms: SpectralTAD (open circles), deDoc (crosses), and Armatus (open squares).}
\label{Fig1}
\end{figure}

Because there is currently no clear ground-truth definition of TADs in Hi-C maps, it is challenging to determine the accuracy of TAD calling algorithms on experimental data. However, TAD calling algorithms can be tested on synthetic data that mimics experimental Hi-C maps. The advantage of synthetic data is that it has a well-defined ground-truth and the noise and sparsity of the data can be tuned. To generate a possible ground truth for experimental Hi-C maps, a consensus manual annotation from multiple experts can be obtained. We can then benchmark TAD calling algorithms on their accuracy compared to the manually annotated experimental data~\cite{bib27}.

Many algorithms have been developed to identify TADs using graph-theoretic, clustering, machine-learning, and image transform techniques~\cite{bib10,bib20,bib21,bib22,bib23,bib24,bib25,bib44,bib45}. In Fig. 1D we compare three state-of-the-art TAD calling algorithms on synthetic data generated by sampling from a negative binomial distribution meant to mimic experimental mouse Hi-C maps. These TAD callers identify different numbers of TADs and in different locations, as expected from previous TAD identification algorithm comparison studies ~\cite{bib26,bib27,bib28,bib29,bib30}. Previous studies have found that on manually annotated GM12878 and hESC Hi-C maps at 50 kb resolution, current TAD calling algorithms rarely exceed a positive predictive value of $40 \%$~\cite{bib27}. On synthetic data for overlapping and nested TADs, these methods mostly obtain a true positive rate of $\lesssim 0.6$~\cite{bib28,bib30}. In addition, most current TAD-calling algorithms impose strong restrictions that limit their ability to call overlapping, nested, and gapped TADs.~\cite{bib26,bib27,bib28,bib29,bib30}. 

In this article, we develop a novel TAD-calling algorithm, KerTAD, that applies gradient and other image operators on Hi-C maps to accentuate and extract their off-diagonal features. We show that KerTAD is more accurate than the current state-of-the-art methods as determined by previous studies ~\cite{bib26,bib28,bib29,bib30} across three categories of Hi-C maps: synthetic maps generated via molecular dynamics simulations of block copolymers; synthetic maps with overlapping and nested TADs sampled from a binomial distribution of intensities; and manually annotated GM12878 maps at 50kb resolution. On all three datasets, KerTAD is the most accurate in terms of TPR while having a negligible false discovery rate (FDR). On synthetic data, our method has an average TPR of $\approx 0.97$ and $\approx 0.95$ on non-nested and nested maps, respectively, and a TPR of $\approx 0.80$ on manually annotated Hi-C maps. In addition, KerTAD is highly resistant to noise and sparsity, achieving a higher TPR at the highest level of noise tested than other methods with no noise. Because KerTAD outperforms every tested method on both manually annotated experimental and synthetic data, KerTAD is likely able to capture the underlying features in experimental Hi-C maps.

This article is organized as follows. In the Materials and methods section, we first describe the preprocessing of the input Hi-C maps and the generation of masks to identify key features of TADs in Hi-C maps. We also define the metrics for sensitivity and false discovery rate for comparing the predictions of KerTAD to ground truth for the synthetic and manually annotated Hi-C maps. We then define the techniques used for generating noise and sparsity in synthetic data. In the results section, we summarize the performance of KerTAD (as well as six other methods) in TAD identification on synthetic and manually annotated Hi-C maps. We also analyze replicate Hi-C maps across four organisms and compare the variation in number and mean size of TADs identified by three TAD identification algorithms. Finally, we discuss how the improved accuracy in TAD identification will enable more robust inferences between the identified TADs and chromatin organization.

\section*{Materials and methods}

The description of the Materials and methods is organized into two sections. In the first section, we explain the new TAD identification algorithm, KerTAD, including the preprocessing steps and the application of masks to identify key features of TADs. In the second section, we discuss the implementation of six other state-of-the-art methods to identify TADs, metrics that we use to quantify the accuracy of the TAD identification methods, and techniques to generate sparse and noisy synthetic data. We describe the motivation and process of manually annotating experimental Hi-C maps, as well as the methods for comparing the accuracy of TAD identification methods on manually annotated experimental data. We finally describe in detail our analysis of the performance of several TAD identification algorithms on replicate non-annotated experimental Hi-C maps across several organisms. 

\subsection*{KerTAD}

KerTAD takes as input a symmetric $N \times N$ matrix, $\mathcal{A}_{ij}$, which gives the frequency of contacts between bins $i$ and $j$ and returns an $M \times 2$  matrix, where each row gives the corner location of one of the $M$ TADs in ${\cal A}_{ij}$. The preprocessing step normalizes $\mathcal{A}_{ij}$ such that ${\cal A}_{ii} \geq {\cal A}_{ij}$ for all $i,j$ and reduces fluctuations in ${\cal A}_{ij}$ while preserving edge features. The method then feeds the preprocessed Hi-C map into two separate pipelines, each of which generates a mask. One pipeline seeks to extract small-scale diffuse point features in the Hi-C map, while the other favors larger scale regions near corner points. The final TADs are given by the intersection of the two masks. 

\subsubsection*{Preprocessing}

There is no standard format or normalization scheme for Hi-C maps ~\cite{bib31,bib32,bib33,bib34,bib35,bib36,bib37}. Because normalization is known to significantly affect TAD-calling performance~\cite{bib31}, we first preprocess $\mathcal{A}_{ij}$ to satisfy the requirements below. First,  we ensure that the diagonal elements of $\mathcal{A}_{ij}$ are the maxima in their respective rows, i.e. $\mathcal{A}_{ii} \geq A_{ij}$. If a given ${\cal A}_{ij} > {\cal A}_{ii}$, we then set ${\mathcal A}_{ii} = {\mathcal A}_{ij}$. This condition is reasonable in the sense that we should expect that local regions of chromatin interact with themselves more than any other region. We then locally row-normalize by re-setting ${\mathcal A}_{ij}$ to $(\mathcal{A}_{ij} - \sum_{j=1}^N \mathcal{A}_{ij}/N)/\sigma_i$, where $\sigma_i$ is the standard deviation of the $i$th row of $\mathcal{A}_{ij}$. This normalization reduces global fluctuations and also perturbs the original $\mathcal {A}_{ij}$ less than other normalization schemes like requiring $\mathcal{A}_{ij}$ to be both row- {\it and} column-normalized. 

Once $\mathcal{A}_{ij}$ meets the normalization conditions, we optionally perform total variation regularization to reduce the local fluctuations in ${\cal A}_{ij}$~\cite{bib38,bib39}. The total variation of $\mathcal{A}_{ij}$ is defined as:
\begin{eqnarray}
\label{eq:tv}
	\ V(\mathcal{A}_{ij}) = \sum_{i=1}^{N} \sum_{j=1}^{N}  |\Delta_y {\cal A}_{ij}| +  
    |\Delta_x {\cal A}_{ij}|, 
\end{eqnarray}
where $\Delta_y {\cal A}_{ij} = \mathcal{A}_{(i+1)j}-\mathcal{A}_{ij}$, $\Delta_x \mathcal{A}_{ij} = \mathcal{A}_{i(j+1)}-\mathcal{A}_{ij}$, and the outside bins of $\mathcal{A}_{ij}$ are given by $\mathcal{A}_{(N+1)j} = \mathcal{A}_{Nj}$, $\mathcal{A}_{i(N+1)} = \mathcal{A}_{iN}$, $\mathcal{A}_{i0} = \mathcal{A}_{i1}$, and $\mathcal{A}_{0j} = \mathcal{A}_{1j}$. This ``anisotropic" form (i.e. the sum of $|\Delta_y {\cal A}_{ij}|$ and   
    $|\Delta_x {\cal A}_{ij}|$) for the total variation accentuates vertical and horizontal features in ${\cal A}_{ij}$~\cite{bib40}.
While spatial variation is a hallmark of TADs, excessive variation outside of TAD boundaries (such as speckle noise) can obscure the signal and make TAD identification challenging. While standard smoothing techniques, like Gaussian blurring, can reduce the total variation, they can remove stark edge features that are essential for identifying TADs. We perform an edge-preserving filtering technique by minimizing the following function:
\begin{eqnarray}
\label{eq:constraint}
	\quad V({X}_{ij}) + \lambda     ||{X}_{ij}-\mathcal{A}_{ij}||_1
\end{eqnarray}
over $X_{ij}$, where $\lambda$ controls the strength of the term that penalizes deviations of ${X}_{ij}$ from $\mathcal{A}_{ij}$ and the $L1$ norm is defined as $||Y_{ij}||_1 = \sum_{i=1}^{N} \sum_{j=1}^{N} |Y_{ij}|$. The minimization of Eq.~\ref{eq:constraint} is performed via the Primal-Dual algorithm ~\cite{bib41}.  We set $\lambda=1$ based on finding the $\lambda$ that maximizes TPR across different total variations for maps at different levels of noise and sparsity (Fig S1B). 
Finally, we filter $\mathcal{A}_{ij}$ with a Gaussian kernel with standard deviation $\sigma= 3\Gamma/2$ and filter size $2\lceil (2\sigma)\rceil +1$, where $\Gamma N^2$ is the number of zero elements in $\mathcal{A}_{ij}$ and $\lceil \cdot \rceil$ is the ceiling function. This Gaussian filtering is performed since extremely sparse Hi-C maps can cause division by zero errors in the KerTAD masks.  

\subsubsection*{Mask for corner point features}

The mask for corner point features is designed to identify locations near the diagonal where there are strong changes in intensity, since these often indicate transitions between TADs, and then to generate a mask of possible corner point combinations in $\mathcal{A}_{ij}$.  We first calculate the discrete partial derivative of $\mathcal{A}_{ij}$. We then feed the row vectors of the partial derivative map into a non-linear function that produces a similarity matrix. The similarity matrix is then filtered by applying a local maximum operator and global threshold, which identifies locations on the diagonal of $\mathcal{A}_{ij}$ where there are sharp local changes. We then use the identified locations on the diagonal to generate a binary mask of every TAD corner point combination, with each diagonal location representing one index of a possible TAD corner point. 
Differential operators in image processing are often represented as convolutions of an image with a kernel that is separable into at least one smoothing filter. Smoothing can reduce noise, but excessive smoothing removes edge features, making it difficult to determine TAD locations. Thus, we implement a low-order partial derivative map with no smoothing filter, $\Delta_y \mathcal{A}_{ij}$, with symmetric boundary conditions.  

Next, we construct a list of row vectors $\{{\vec v}_1,..,{\vec v}_N\}$, where ${\vec v}_i$ is the $i$th row of $\Delta_y \mathcal{A}_{ij}$. We then construct a similarity matrix, ${\cal S}_{ij}$, 
\begin{eqnarray}
\label{eq:schemeP}
\mathcal{S}_{ij} = \left(\max(\vec{v}_i)-\min(\vec{v}_i)+\max(\vec{v}_j)-\min(\vec{v}_j)\right) ||\vec{v}_i||_1 ||\vec{v}_j||_1,
\end{eqnarray}
and $\max ({\vec v}_i)$ and $\min ({\vec v}_i)$ return the maximum and minimum components of ${\vec v}_i$, respectively. Finally, we define the $N \times N$ binary mask of point features, $\mathcal{M}_{ij}$, as  follows: for every $i,j \text{ such that } i < j$,  $\mathcal{M}_{ij}=1$ if and only if $\mathcal{S}_{ii}$ and $\mathcal{S}_{jj}$ are both local maxima in their respective $3\times 3$ local neighborhoods \emph{and} $\mathcal{S}_{ii}, \mathcal{S}_{jj} \geq \Omega$, where $\Omega$ is the global threshold determined using the triangle algorithm\cite{bib43} on $\mathcal{S}_{ij}$.  Fig. 2 illustrates the several intermediate steps and maps to transform an input Hi-C map, $\mathcal{A}_{ij}$, into $\mathcal{M}_{ij}$.

\begin{figure}[!h] 
\includegraphics[width=\textwidth]{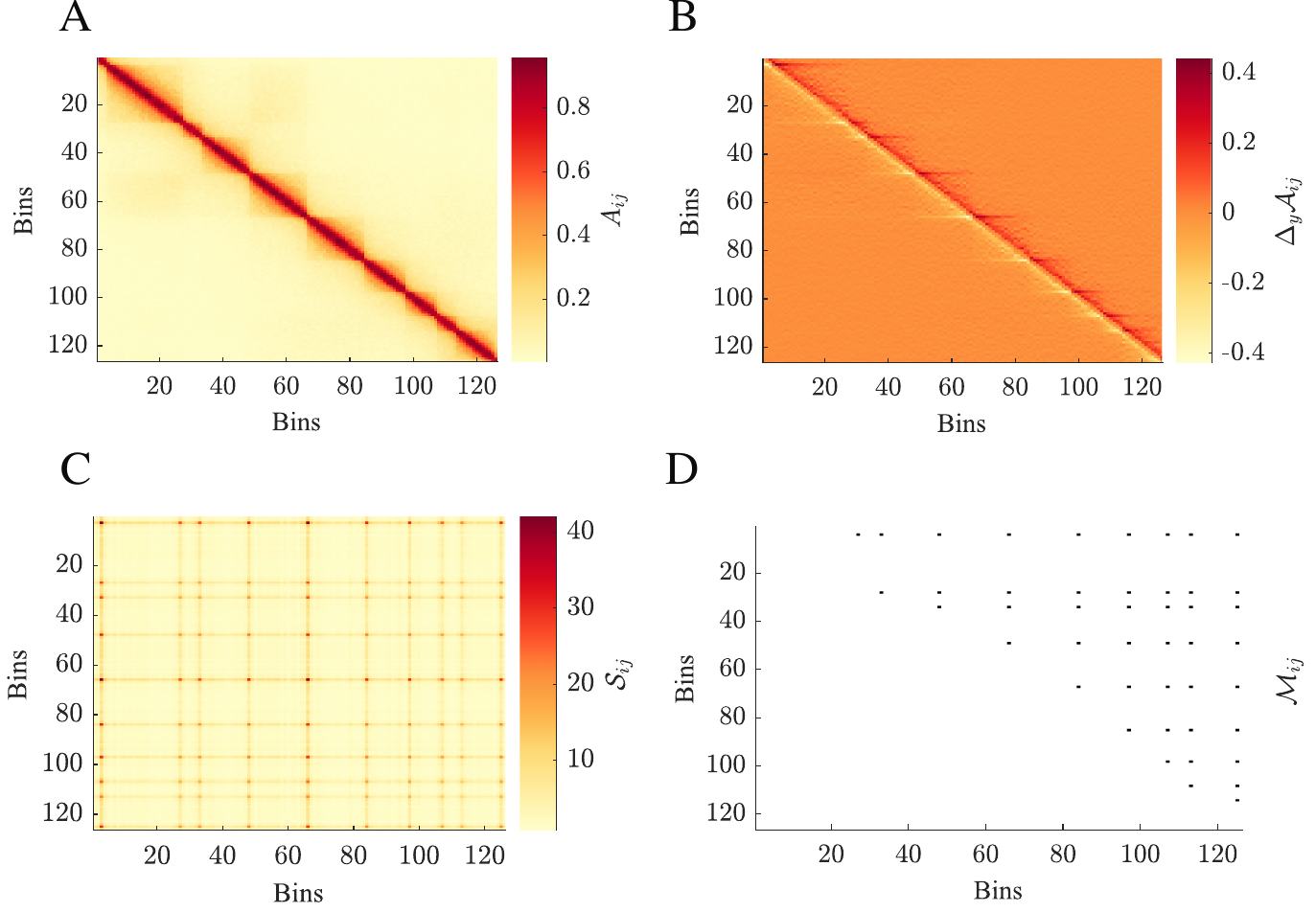}
\caption{{\bf Illustration of the four steps in constructing the point feature binary mask.}
A: We start with a Hi-C map ${\cal A}_{ij}$ (with bins $i$ and $j$ labelled from $1$ to $125$). B: We first calculate the discrete partial derivative, $\Delta_y A_{ij}$. C: We then construct $\mathcal{S}_{ij}$ from a nonlinear function of the pairs of row vectors of $\Delta_y {\cal A}_{ij}$. D: The binary mask ${\cal M}_{ij}$ is obtained by combining a local maximum filter with binary thresholding of $\mathcal{S}_{ij}$. If ${\cal M}_{ij} = 1$ (black squares), $\mathcal{S}_{ii}$ and $\mathcal{S}_{jj}$ are both local maxima in their $3 \times 3$ windows and above the threshold set by the triangle method on $\mathcal{S}_{ij}$.}
\label{Fig2}
\end{figure}

\subsubsection*{Mask for corner regions}

While the previous mask captured $\emph{point}$ features of TADs spread throughout the Hi-C map, we also need a mask to identify the specific corner $\emph{regions}$ near the diagonal in ${\cal A}_{ij}$. As before, we calculate an image derivative, this time $\Delta_x \mathcal{A}_{ij}$, using periodic boundary conditions. For $i<j$, if $\Delta_x \mathcal{A}_{ij}>0$ then $\Delta_x \mathcal{A}_{ij}$ is set to $0$ and for $i>j$ if $\Delta_x \mathcal{A}_{ij}<0$ then $\Delta_x \mathcal{A}_{ij}$ is set to $0$. We then calculate
\begin{eqnarray}
\mathcal{P}_{ij} = \sum_{k=1}^N \left( \Delta_x \mathcal{A}_{ik} \Delta_x \mathcal{A}^T_{kj} - \Delta_x \mathcal{A}^T_{ik} \Delta_x \mathcal{A}_{kj}\right).
\end{eqnarray}
$\mathcal{P}_{ij}$ has several important features. First, TAD corners and edges are maxima of $\mathcal{P}_{ij}$ in their local neighborhood as shown in Fig. 3. The diagonal elements of $\mathcal{P}_{ij}$ that correspond to TAD corner points (i.e. if ${\cal A}_{ij}$ is the corner point of a TAD, the corresponding points in $\mathcal{P}_{ij}$ are $\mathcal{P}_{ii}$ and $\mathcal{P}_{jj}$) are strongly negative minima in their neighborhood. Taking advantage of both of these facts, we construct the final binary mask $\mathcal{M}'_{ij}$: 

\begin{eqnarray}
\mathcal{M}'_{ij}=
    \begin{cases}
        1 & \text{if } (-\mathcal{P}_{ij} (\mathcal{P}_{ii}+\mathcal{P}_{jj}))\geq \Omega\\
        0 & \text{otherwise},
    \end{cases}
\end{eqnarray}
where $\Omega$ is the threshold determined by the triangle method on the matrix, $\mathcal{C}_{ij} = -\mathcal{P}_{ij} (\mathcal{P}_{ii}+\mathcal{P}_{jj})$.

\begin{figure}[!h] 
\includegraphics[width=\textwidth]{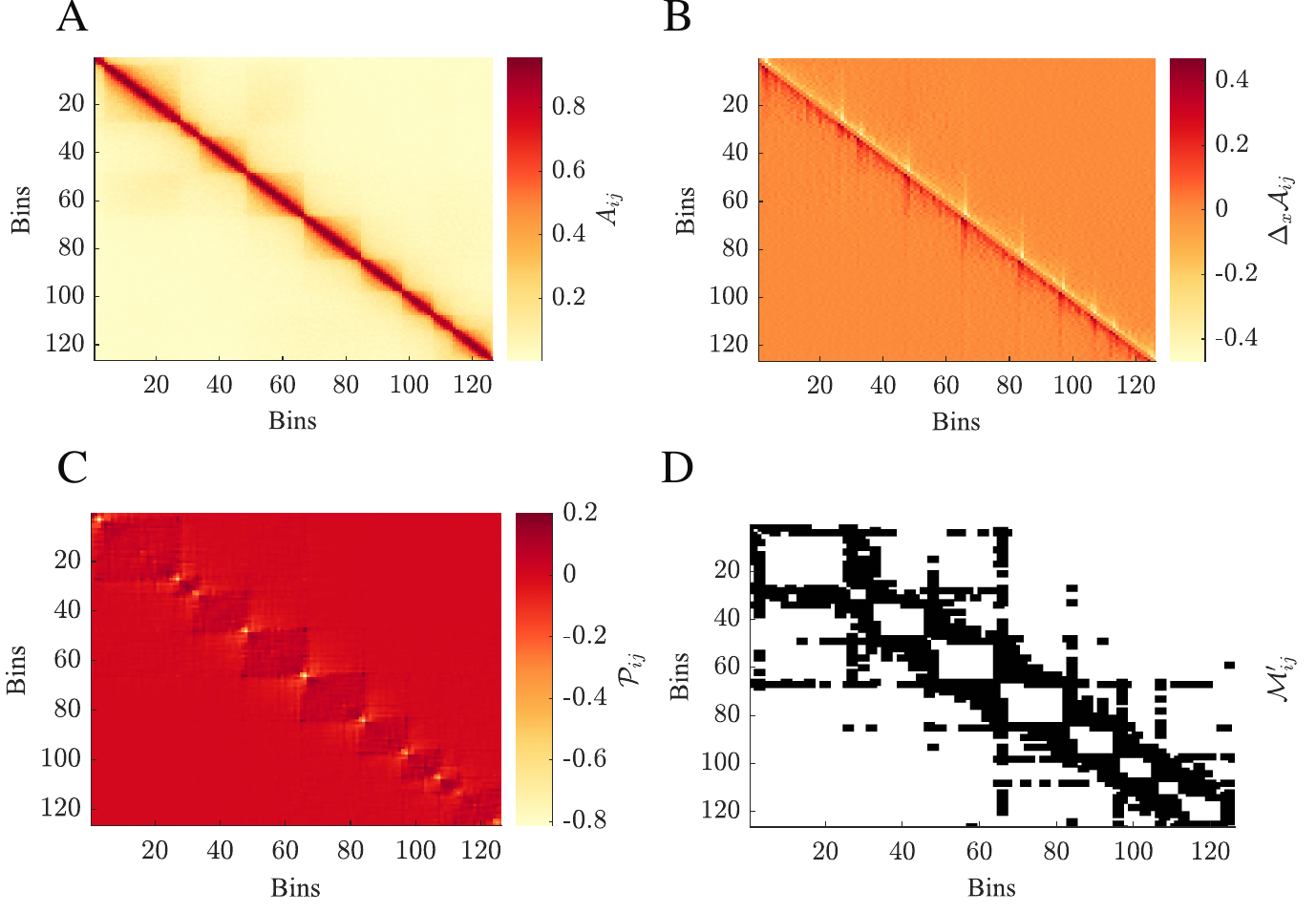}
\caption{{\bf Illustration of the steps used to construct the mask ${\cal M'}_{ij}$ for identifying corner regions in Hi-C maps.}
A: We start with the same input Hi-C map ${\cal A}_{ij}$ as in Fig 2. B: We first calculate the discrete partial derivatives, $\Delta_x A_{ij}$. C: We then calculate $\mathcal{P}_{ij}$ from $\Delta_x A_{ij}$. D: We  obtain the final binary mask ${\cal M'}_{ij}$ after applying a global threshold on -$\mathcal{P}_{ij}(\mathcal{P}_{ii}+\mathcal{P}_{jj})$.}
\label{Fig3}
\end{figure}

\subsubsection*{Final Mask}

After constructing both masks, we take the element-wise product of $\mathcal{M}$ and $\mathcal{M'}$ to obtain the final binary mask,  $\mathcal{{B}}_{ij} = \mathcal{M}_{ij} \mathcal{M}'_{ij}$. Each nonzero element of $\mathcal{{B}}$ represents a predicted TAD corner point. For the final output, KerTAD converts $\mathcal{{B}}$ to a $2$ column list where each row represents the start and end index of a TAD corner point. 
KerTAD, while not requiring any explicit user provided parameters, has several optional parameters to improve flexibility for the user.   First, $\phi$ is a binary variable such that when $\phi=0$ (by default), total variation regularization (TVR) is not performed. TVR can be computationally expensive and for Hi-C maps with low total variation, TVR is not necessary. We also set $\kappa$ as the maximum number of TADs that can be identified per row. Lastly, when the binary parameter $\gamma = 1$, $A_{ij}$ is broken into smaller maps to process each separately. Splitting the Hi-C maps is useful for large and heterogeneous Hi-C maps where different regions have significantly different coverage and local intensity. Unless otherwise noted, we use the default settings for calculations throughout the manuscript, i.e. $\phi, \gamma=0$, and $\kappa=3$ .

\subsection*{Benchmarks}

When determining the accuracy of TAD identification methods, we first categorize the Hi-C maps into two types: synthetic and experimental Hi-C maps. For synthetic Hi-C maps, we also distinguish between ``simple" and ``complex" Hi-C maps. For simple Hi-C maps, each element on the diagonal of $\mathcal{A}_{ij}$ must belong to one and only one TAD. This condition implies that i) ${\cal A}_{ij}$ has no nested or overlapping TADs and ii) ${\cal A}_{ij}$ has no gaps between TADs. Thus, in a simple Hi-C map, if a TAD is identified over a set of diagonal elements, e.g. from $A_{ii}$ to $A_{jj}$, there are no other TADs within that set and the next TAD must start at $A_{(j+1)(j+1)}$. Complex Hi-C maps are defined as any Hi-C map ${\cal A}_{ij}$ that is not simple, i.e. ${\cal A}_{ij}$ has either nested, overlapping, or gapped TADs. A nested TAD is a TAD with its corner point located at ${\cal A}_{ij}$ (where $j>i$) while there exists another TAD corner at ${\cal A}_{kl}$  (where $l>k$), where $k\leq i \text{ and } j \leq l$. An overlapping TAD has a corner at ${\cal A}_{ij}$ (where $j>i$) and another TAD corner at ${\cal A}_{kl}$ (where $l>k$), where $k<i \text{ and } i<l<j $  or $ i<k<j \text{ and } j<l$.  A Hi-C map possesses a gapped TAD if there exists an element on the diagonal, ${\cal A}_{ii}$, that does not belong to any TAD.

We analyze the performance of TAD identification algorithms on simple and complex synthetic Hi-C maps separately. Many TAD identification algorithms assume that the input Hi-C maps are simple. This additional information provides constraints on the locations of TADs, which can lead to enhanced accuracy for these algorithms. However, the additional constraints do not improve TAD prediction in manually annotated experimental Hi-C maps, as most experimental Hi-C maps are not simple. In previous work comparing the performance of TAD identification algorithms, the top performers on simple and complex synthetic maps were different~\cite{bib28,bib30}. In the Results section, we show that KerTAD is highly accurate in identifying TADs in both simple and complex Hi-C maps, while not presupposing that a given Hi-C map is simple or complex.

\subsubsection*{Simple Synthetic Hi-C Maps}
To compare the performance of different TAD identification algorithms for simple, synthetic Hi-C maps, we consider $100$ Hi-C maps generated by molecular dynamics (MD) simulations of block copolymers from previous studies~\cite{bib42}. In these MD simulations, chromatin is modeled as a bead-spring polymer with non-bonded, purely repulsive interactions to prevent bead overlaps, non-specific short-ranged attractive interactions between bead pairs to induce compaction, and specific short-ranged attractive interactions between bead pairs to mimic TADs that occur in specific epigenomic profiles. 

From previous studies ~\cite{bib26, bib27, bib28, bib29, bib30} we select the top performing TAD identification algorithms for simple, synthetic maps. Namely, we compare KerTAD with TopDom~\cite{bib25}, HICSeg~\cite{bib44}, and CHDF~\cite{bib45}. We perform TAD identification on the set of $100$ simple, synthetic Hi-C maps discussed above. (Note that TopDom, HICSeg, and CHDF do not identify nested or overlapping TADs.) For TopDom we count the "domain" predictions and set the window size to $5$ as done in previous work \cite{bib28,bib30} for the same synthetic Hi-C maps. Again following previous work \cite{bib26,bib27,bib28,bib30}, we set the max TAD size parameter for CHDF to 50 and for HICSeg we use the "G" distribution. When comparing TAD predictions from KerTAD to those for the other algorithms on the simple, synthetic Hi-C maps, we impose a further restriction on our identified TADs. Since KerTAD can identify nested and overlapping TADs, it has more chances to identify correct TADs compared to methods that are unable to call nested and overlapping TADs. Thus, we set $\kappa=1$, considering only the innermost TAD corners with the smallest distance to the diagonal. 

\subsubsection*{Complex Synthetic Hi-C Maps}

For generating complex, synthetic Hi-C maps, we use a variation of a previously developed procedure ~\cite{bib26, bib46} that mimics mouse embryonic stem cells by sampling from a negative binomial distribution of Bernoulli trials, where successful trials represent contacts between chromatin loci. The distribution is characterized by a location-dependent variance $\sigma^2_{ij} = \mu_{ij} + r \mu_{ij} ^2$ (with dispersion factor $r=0.01$) and mean $\mu_{ij} = \langle {\cal A}_{ij} \rangle$.  The location-dependent mean is defined by 
\begin{eqnarray}
\mu_{ij} = K_d \delta_{ij}  +  \theta_{ij} K_t(i-j+1)^c + \mathcal{N}_{noise},
\end{eqnarray}
where $\delta_{ij}$ is the Kronecker-delta, $K_d$ gives $\langle {\cal A}_{ii} \rangle$, $K_t$ and $c$ are parameters that control the power-law decay of $\langle {\cal A}_{ij}\rangle$ away from the diagonal. ( $K_d=35$, $K_t = 28$ and $c= -0.69$ were selected to match $\langle {\cal A}_{ij}\rangle$ in chromosome five in IMR90 replicate B.) $\theta_{ij}=1$ when $\mathcal{A}_{ij}$ is inside of a TAD (excluding diagonal elements) and $0$ otherwise.  TAD boundary lengths are selected randomly from a uniform distribution with widths from $5$ to $20$ bins (where each bin represents $40$ kb). We then remove randomly selected TADs from this list and fill in the gaps with larger overlapping and nested TADs. $\mathcal{N}_{noise}$ is a random variable that mimics weak and non-specific ligation events by sampling (with replacement) a fraction of  randomly selected elements of $\mathcal{A}_{ij}$ and adding a  constant, $K_{\text{noise}}$ (we set $K_{\text{noise}}=5$).  The likelihood that an element of $\mathcal{A}_{ij}$ receives a noise impulse scales with $(i-j+1)^c$.

We generate $100$ complex, synthetic Hi-C maps using this protocol with $\mathcal{N}_{noise} =0$, where each Hi-C map has on average $150$ TADs. From previous studies~\cite{bib27,bib28,bib29,bib30} we select the top performing TAD callers on similar datasets of complex, synthetic Hi-C maps. We compare KerTAD with deDoc~\cite{bib22}, Armatus ~\cite{bib20}, and SpectralTAD ~\cite{bib23}. As before, we follow the default or recommended parameters for each algorithm. For Armatus we set g=0.05 and s=0.05 \cite{bib26}, for SpectralTAD we use levels=2, and for deDoc we use both the dedoc(M) and dedoc(E) predictions, removing duplicates. The accuracy of TAD identification was determined for these three methods, along with KerTAD, for each complex, synthetic Hi-C map.

\subsubsection*{Noise and Sparsity}

To test the robustness of the TAD identification algorithms, we compare TAD predictions for two sets of new complex, synthetic Hi-C maps with varying levels of added noise and sparsity. In the first set, we generate $10$ complex Hi-C maps with $\mathcal{N}_{noise}=0$ (as previously described) and for each, construct an additional $20$ Hi-C maps, with varying levels of noise (totalling 210 total Hi-C maps). Because many TAD identification algorithms only accept integer counts, we do not use additive Gaussian noise. Instead, we randomly sample $\mathcal{A}_{ij}$ (with replacement) and add a constant additive impulse, $K_{\text{noise}}=5$, as described previously for $\mathcal{N}_{noise}$. The noise is parameterized by $\chi$, which represents the number of added impulses divided by the number of elements of $\mathcal{A}_{ij}$. To generate the noisy maps, we increase $\chi$ in increments of $0.05$ starting from $0$ to $1$ . For the second set, we perform the same procedure but instead add sparsity to ${\cal A}_{ij}$ by setting random elements of ${\cal A}_{ij}$ equal to 0. Sparsity is parameterized by $\xi$, which is the fraction of elements of ${\cal A}_{ij}$ that are set to zero compared to the total number of elements. We generate $200$ sparse maps by increasing $\xi$ in increments of $0.05$ starting from $0$ to $0.95$ ($\xi=1$ would mean a map of only 0s). 

\subsubsection*{Experimental Maps}

To obtain ground truth for experimental Hi-C maps, we follow the previous manual annotations performed on Hi-C maps for the GM12878 cell line at $50$ kb resolution for the $40$–$45$ Mb regions of $10$ different chromosomes (chromosomes $2$, $3$, $4$, $5$, $6$, $7$, $12$, $18$, $20$, and $22$)~\cite{bib27}. In the original annotations, "any identifiable TAD structure"  was annotated and the positive predictive value (PPV) of the identified TADs was calculated for seven TAD identification algorithms\cite{bib27}. However, calculating PPV does not penalize TAD callers that miss "obvious" TADs and even TPR may be inappropriate for gauging TAD prediction accuracy if the annotations are forgiving enough. In addition, likely due to differences in the pipeline or visualization, we found that many of the original annotations were displaced or pointed at no features or structures. Thus, using the original annotations as a guide, we keep the most "obvious" TADs and then calculate TPR to capture the accuracy of the TAD identification methods. Because the annotations are not meant to be exhaustive, we do not calculate FDR. Because the experimental Hi-C maps are complex, we use deDoc, Armatus, and SpectralTAD, as well as KerTAD, to identify TADs in the manually annotated GM12878 Hi-C maps. For the input maps to each TAD caller, we used the cutout sections of the genome except for Armatus which returned no TADs with the smaller map (a previously described bug) and for which we used the full intrachromosomal map as input. 

For experimental Hi-C maps without manual annotations, we evaluate in situ Hi-C maps for four organisms: fruit fly S2 cells ~\cite{bib47} (4DN accession code: 4DNESFOADERB), zebrafish embryos ~\cite{bib48} (4DN accession code: 4DNESV5PGOUC), mouse CH12.LX cells ~\cite{bib17} (4DN accession code: 4DNESK95HVFB), and human HCT-116 cells ~\cite{bib50} (4DN accession code: 4DNES3QAGOZZ). All Hi-C maps were obtained from the 4DN data portal and the {\tt .pairs} files for each biological and technical replicate were converted to {\tt .cool} files and then intrachromosomal Hi-C maps at $50$ kb resolution were extracted using Cooler ~\cite{bib50}. For zebrafish Hi-C maps, we analyzed three biological replicates with one technical replicate for each biological replicate. For fruit fly Hi-C maps, we also analyzed three biological replicates with one technical replicate each. For mouse Hi-C maps, we used three biological replicates with $11$, $2$, and $2$ technical replicates. For human Hi-C maps, we analyzed six biological replicates with $3$, $4$, $2$, $3$, $2$, and $2$ technical replicates. For each Hi-C map, we perform TAD identification using KerTAD and the top performers in TPR for the simple and complex Hi-C map categories: TopDom and deDoc. For TopDom we used a window size of $10$ following the recommendation for 50kb resolution from previous work \cite{bib27}. Because TopDom threw an error for chromosome Y of biological replicate 2 for fruit fly, we do not include that Hi-C map in our analysis for TopDom. We calculate the total number of identified TADs by summing the number of predicted TADs for each intrachromosomal map for each replicate. We also calculate the mean size of the identified TADs for each intrachromosomal map. We characterize the distribution of the number of TADs and mean sizes of TADs over replicates for each organism by calculating the median, maximum, and minimum values.

\subsubsection*{Metrics}

We apply each TAD identification algorithm to each synthetic or manually annotated experimental Hi-C map and compare the lists of identified TADs to ground truth. For a predicted TAD corner point located at $\mathcal{A}_{ij}$, we call it a "true positive" if and only if there is a ground truth TAD with the same corner point coordinates. We calculate two metrics for each synthetic and experimental Hi-C map for every algorithm: ${\rm TPR}=p/{\cal G}$ and ${\rm FDR}=({\cal T}-p)/{\cal T}$, where $p$ is the number of true positives, $\mathcal{G}$ is the total number of ground truth TADs, and $\mathcal{T}$ is the total number of TADs predicted. In manually annotated experimental Hi-C maps, since the TAD corners are often difficult to define, a ``true positive" is counted as long as the ground truth coordinate is one of the coordinates in the $3 \times 3$ square centered around the predicted TAD corner point.

\section*{Results}

In this section, we compare the performance of KerTAD against current state-of-the-art TAD identification methods using two metrics: the ability to reliably identify ground truth TADs (TPR) and the ability to avoid predicting incorrect TADs (FDR). We compare the accuracy of seven different methods on two sets of synthetic Hi-C maps: a set of simple Hi-C maps obtained from MD simulations of block copolymers and a set of complex Hi-C maps generated by sampling a negative binomial distribution. We also calculate TPR and FDR for the same TAD identification algorithms on manually annotated Hi-C maps from the GM12878 cell line. Finally, we calculate the number and size of TADs obtained using each algorithm on in-situ experimental Hi-C maps for four organisms: mouse, human, fruit fly, and zebrafish. 

On the $100$ simple, synthetic Hi-C maps, our method gives the highest median ${\rm TPR} \approx 0.99$ and the lowest median ${\rm FDR} \approx 0.02$ of all surveyed methods (Fig. 4A). The next best performing algorithm, TopDom, had a comparable median ${\rm TPR} \approx 0.94$ and median ${\rm FDR} \approx 0.03$, but TopDom yields a significantly larger variance with a minimum ${\rm TPR} \approx 0.65$ compared to $\approx 0.88$ for our method. In Fig. 4A, we also show that the other TAD identification algorithms, CHDF and HiCSeg, performed poorly on the simple, synthetic Hi-C maps with a median ${\rm TPR} < 0.6$ and median ${\rm FDR} > 0.2$. (Note that the median ${\rm FDR} \approx 0.7$ for CHDF was larger than its median ${\rm TPR} \approx 0.5$.) In previous work, ~\cite{bib28,bib30} CHDF was reported to perform very well on this synthetic dataset (hence why it was selected for comparison), scoring a mean ${\rm TPR} \approx 0.965$ and ${\rm FDR} \approx 0.381$ . Even granting these scores, KerTAD still outperforms CHDF in both TPR and FDR. In fact, KerTAD scores a higher mean and minimum TPR than all $27$ surveyed TAD callers in previous works \cite{bib28,bib30}. Furthermore, when running our method on simple, synthetic Hi-C maps, we did not allow it to call nested or overlapping TADs. Without this restriction, the median TPR was even greater than $0.99$, while maintaining small median ${\rm FDR}$.

\begin{figure}[!h]
\includegraphics[width=\textwidth]{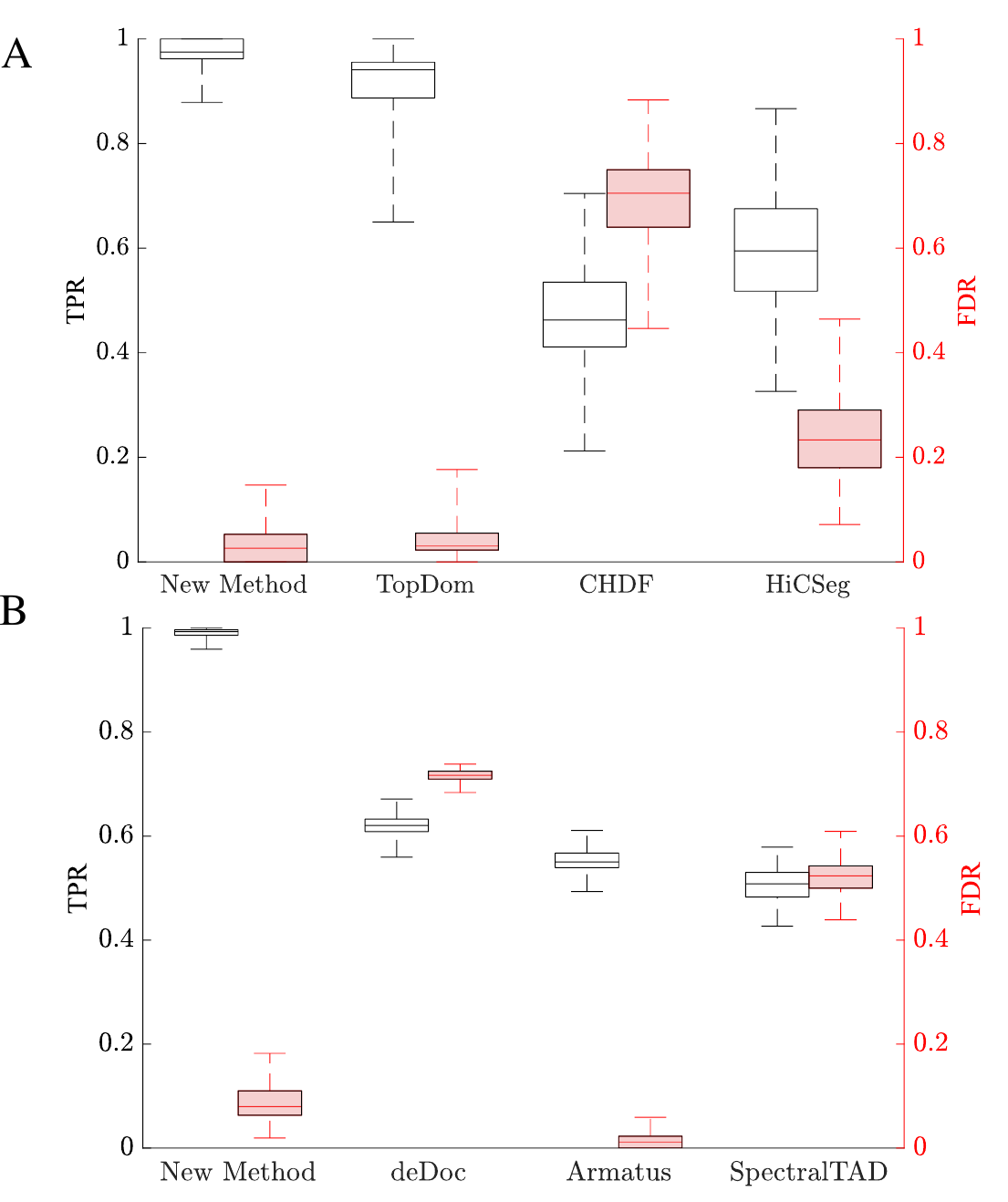}
\caption{{\bf TPR and FDR on simple and complex synthetic Hi-C maps.}
A: Box plots of TPR (black; left axis) and FDR (red; right axis) calculated by comparing the ground truth TADs from $100$ simple, synthetic Hi-C maps generated by MD simulations of block copolymers~\cite{bib42} and TADs predicted by the new method, TopDom, CHDF, and HiCSeg. The box edges represent the $25$th and $75$th percentiles in TPR/FDR, and the central line in each box indicates the median. The error bars represent the maximum and minimum TPR or FDR.  B: Box plots for TPR (black) and FDR (red) for $100$ complex, synthetic Hi-C maps that mimic mouse embryonic stem cells by sampling from a negative binomial distribution~\cite{bib26,bib28}. We show the TPR and FDR for the new method, deDoc, Armatus, and SpectralTAD.}
\label{Fig4}
\end{figure}

For the $100$ complex, synthetic Hi-C maps, the differences in the median TPR between the new method and the other tested algorithms are more pronounced, as shown in Fig. 4B. The new method obtains a median ${\rm TPR} \approx 0.98$, while the next best TAD identification method, deDoc, on complex synthetic maps only had a median ${\rm TPR} \approx 0.65$. The remaining algorithms, Armatus and SpectralTAD, were roughly comparable in TPR performance with deDoc. For FDR, Armatus performed the best (median 0.01) followed by KerTAD (median 0.08). DeDoc and SpectralTAD had significantly higher FDRs with both greater than $0.45$.

We also studied the impact of impulse noise on the calculations of TPR and FDR on complex, synthetic Hi-C maps. We find that our method is highly resistant to noise. In Fig. 5A, we show that the mean TPR decays slowly with increasing $\chi$, i.e. the mean ${\rm TPR} > 0.8$ across all tested values of $\chi$. In contrast, none of the other tested algorithms achieve a mean TPR of $0.70$ or greater at any $\chi$. 

\begin{figure}[!h]
\includegraphics[width=\textwidth]{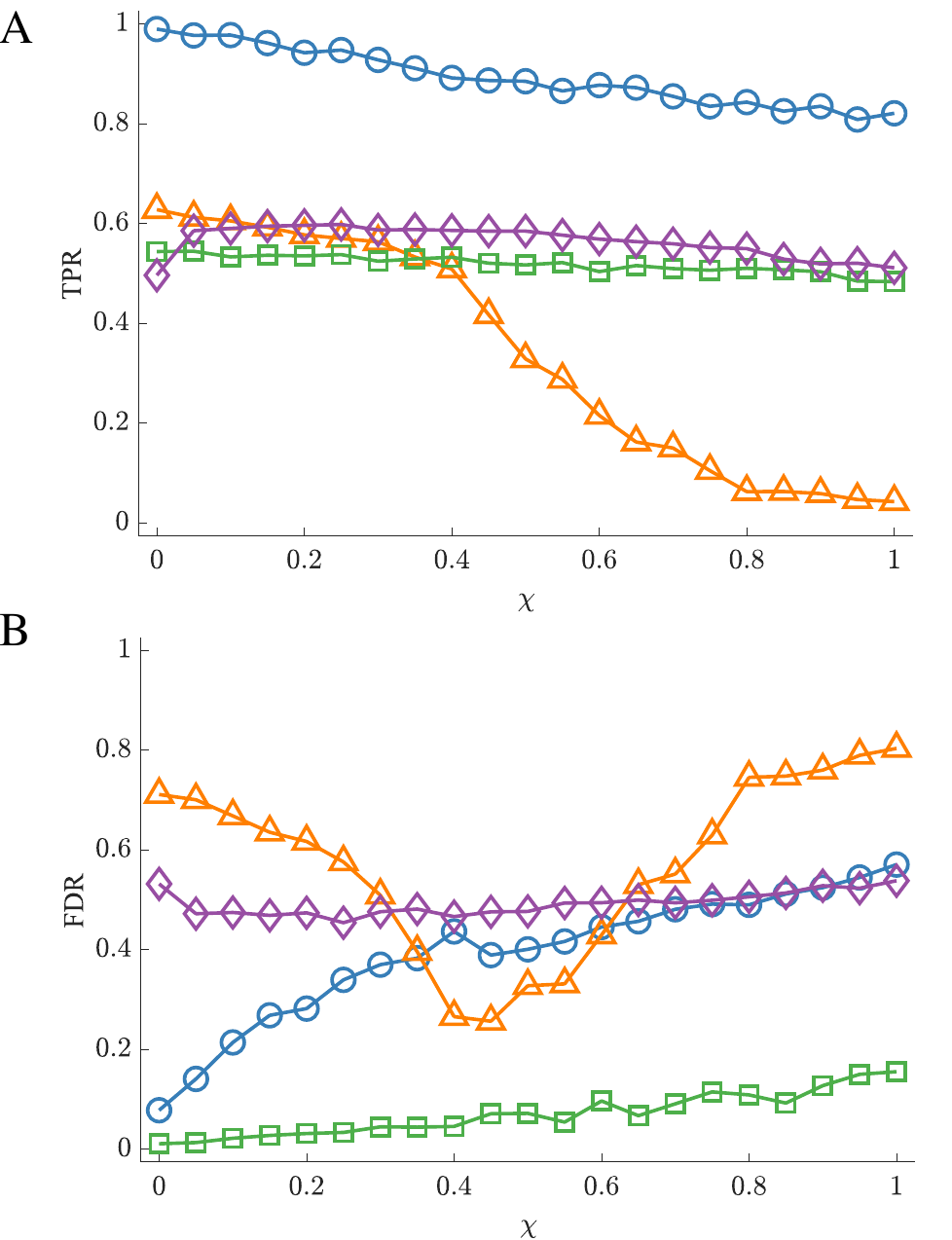}
\caption{{\bf TPR and FDR on Hi-C maps with added noise.}
A: TPR averaged over $210$ complex synthetic Hi-C maps plotted versus $\chi$. We calculate TPR by comparing the ground truth of the synthetic Hi-C maps with the predicted TADs for KerTAD (blue circles), deDoc (orange triangles), Armatus (green squares), and SpectralTAD (purple diamonds). B: FDR plotted versus $\chi$ for the same data in A.}
\label{Fig5}
\end{figure}

In addition, we investigated the effect of sparsity on the ability of TAD identification algorithms to predict TAD locations. To incorporate sparsity, we modify complex synthetic maps by randomly selecting elements in ${\cal A}_{ij}$ and replacing them with $0$. In Fig. 6A, we show that our method achieves a higher mean TPR at almost every $\xi$ than all other tested TAD identification algorithms. We find that the mean TPR for KerTAD is significantly higher for the majority of $\xi$ values tested; for example, our method achieves a higher mean TPR at $\xi =0.5$ than the second best algorithm, deDoc, at $\xi =0$. The mean FDR for our method also grows more slowly compared to the other tested algorithms, only passing a mean FDR of $0.5$ at large sparsity, $\xi > 0.6$. (See Fig. 6B.).  SpectralTAD threw errors with large values of $\xi$ and returned no predicted TADs (for these maps we set ${\rm TPR}=0$ and ${\rm FDR}=1$).

\begin{figure}[!h]
\includegraphics[width=\textwidth]{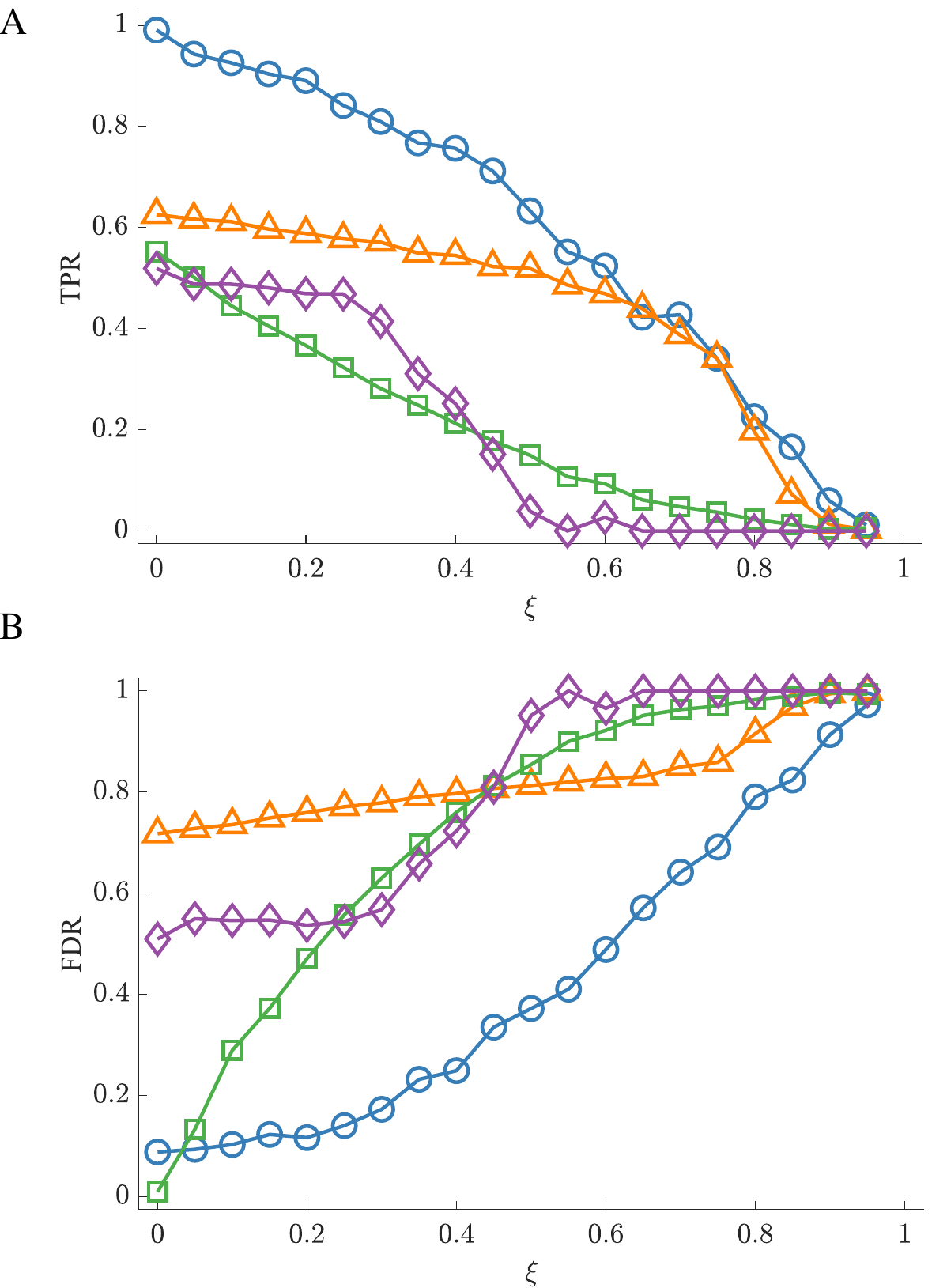}
\caption{{\bf TPR and FDR on Hi-C maps with added sparsity.}
A: TPR averaged over $200$ complex synthetic Hi-C maps plotted versus the sparsity fraction $\xi$. We calculate TPR by comparing the ground truth of the synthetic Hi-C maps with the predicted TADs for the KerTAD (blue circles), deDoc (orange triangles), Armatus (green squares), and SpectralTAD (purple diamonds). B: FDR plotted versus $\xi$ for the same data in A.}
\label{Fig6}
\end{figure}

In addition to assessing the performance of TAD identification algorithms on synthetic Hi-C maps, we also determined their performance on manually annotated Hi-C maps from the GM12878 cell line. We calculated TPR and FDR averaged over 10 chromosomes (chromosomes $2$, $3$, $4$, $5$, $6$, $7$, $12$, $18$, $20$, and $22$) by treating the manual annotations as the ground truth. We show in Fig. 7 that our new method achieves a median TPR of nearly $0.80$, while the next best performer, deDoc, obtains a median TPR of only $\sim 0.4$. When using the original annotations, we also found that KerTAD outperformed the other techniques by a factor of $\approx 2$ (KerTAD had a TPR of $0.4$ while the next best, deDoc, had a TPR of $0.2$). However, we were unable to precisely match the maps the original annotations used, with many annotated TADs pointing to no visible structure and hence the original annotation TPRs are likely not very meaningful.

\begin{figure}[!h]
\includegraphics[width=\textwidth]{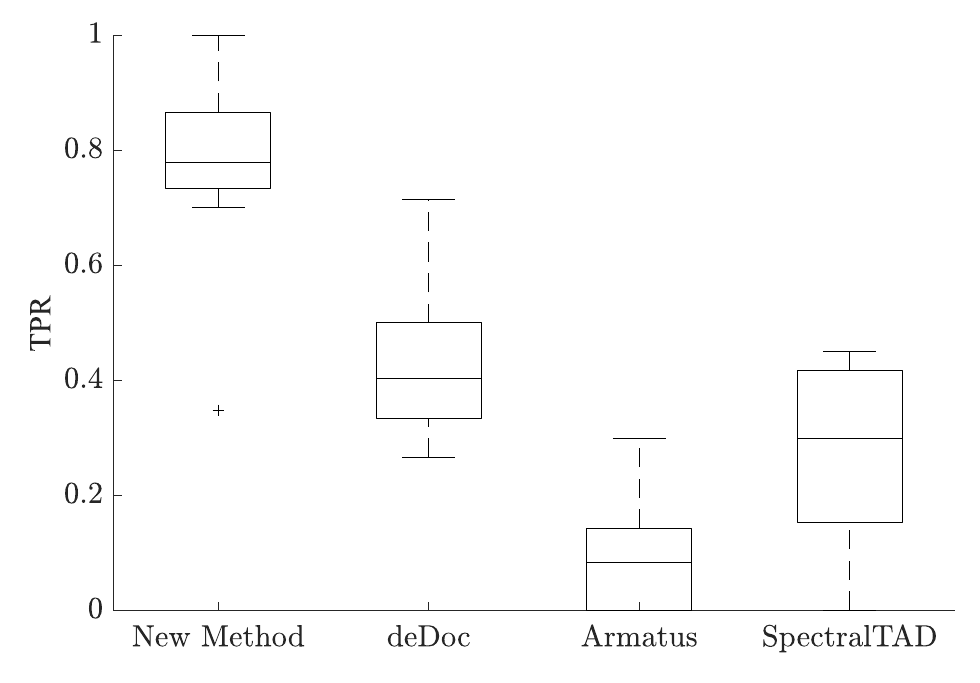}
\caption{{\bf TPR on manually annotated Hi-C maps}
A: Box plots of TPR calculated by comparing the ground truth TADs from ten manually annotated GM12878 Hi-C maps to those predicted by KerTAD, deDoc, Armatus, and SpectralTAD. The box edges represent the $25$th and $75$th percentiles in TPR and the central line in each box indicates the median. The error bars represent the maximum and minimum values of TPR that are not outliers. The "+" represents an outlier.}
\label{Fig7}
\end{figure}

Our new TAD identification method achieves a higher TPR and lower FDR on both simple and complex synthetic Hi-C maps, as well as on manually annotated experimental Hi-C maps. (See Fig. 9A-C for Hi-C maps with superimposed TAD predictions.) Additionally, our new method achieves and maintains the highest TPR in Hi-C maps with added noise and sparsity. Based on these results, we suggest that our method will have the highest accuracy of TAD identification on non-annotated experimental Hi-C maps. We compare the TAD predictions for the top-performing algorithms on synthetic and manually annotated Hi-C maps on non-annotated experimental HI-C maps for four organisms: zebrafish, fruit fly, mouse, and human. In Fig. 8A, we find that deDoc, TopDom, and our method predict different median total numbers of TADs (over the intrachromosomal Hi-C maps for all technical and biological replicates). For example, deDoc gives a median of $3370$ TADs for zebrafish, while TopDom predicts roughly a factor of three fewer TADs. For zebrafish and fruit fly, we find that the fluctuations in the number of predicted TADs (given by the difference in the maximum and minimum values) over replicates for each TAD identification algorithm is smaller than the range in the median predictions between algorithms. Among the TAD identification methods tested, TopDom and our method have comparable variations in the number of TADs among replicates, while deDoc showed larger variations, especially for the human Hi-C maps. In Fig. 8B, we show the predictions of the mean size of TADs identified by each algorithm. For the mouse and fruit fly Hi-C maps, we find small variations among the methods on the mean size of TADs, while for zebrafish and human Hi-C maps there are large differences in the TAD sizes. For human Hi-C maps, our method and TopDom predict similar mean sizes for TADs ( $0.8$-$1.2$~${\rm Mb}$), while deDoc shows large fluctuations in the sizes of TADs among replicates. (Note that the fluctuations in the TAD sizes over replicates obtained from our method and TopDom are comparable.). In Fig. 9D, we show a non-annotated human lymphoblastoid Hi-C map with superimposed TAD predictions from KerTAD, deDoc, and TopDom. While there are some TADs for which all methods agree, we find large variability in the locations and number of predicted TADs. 

\begin{figure}[!h]
\includegraphics[width=\textwidth]{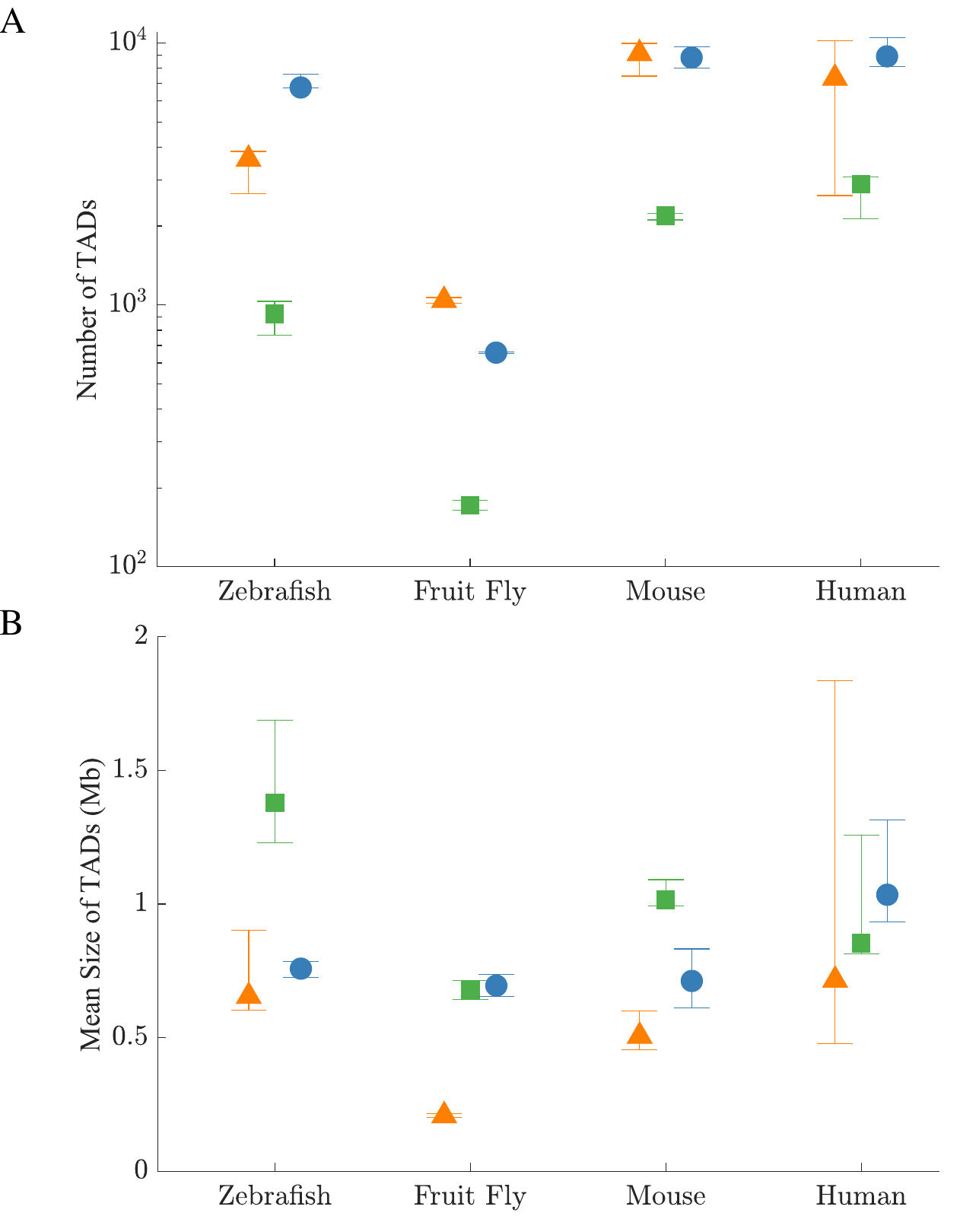}
\caption{{\bf Number and size of TADs predicted across Hi-C map replicates from four organisms.}
A: The number of TADs predicted from whole-genome in situ Hi-C data for technical and biological replicates of four organisms (fruit fly, human, mouse, and zebrafish) using the KerTAD (blue circles), deDoc (orange triangles), and TopDom (green squares). The symbols indicate the median number of TADs and the error bars indicate the maximum and minimum values over replicates. B: Average size of the TADs in Mb for the same organisms, set of replicates, and TAD identification algorithms in A.}
\label{Fig8}
\end{figure}

\begin{figure}[!h]
\includegraphics[width=\textwidth]{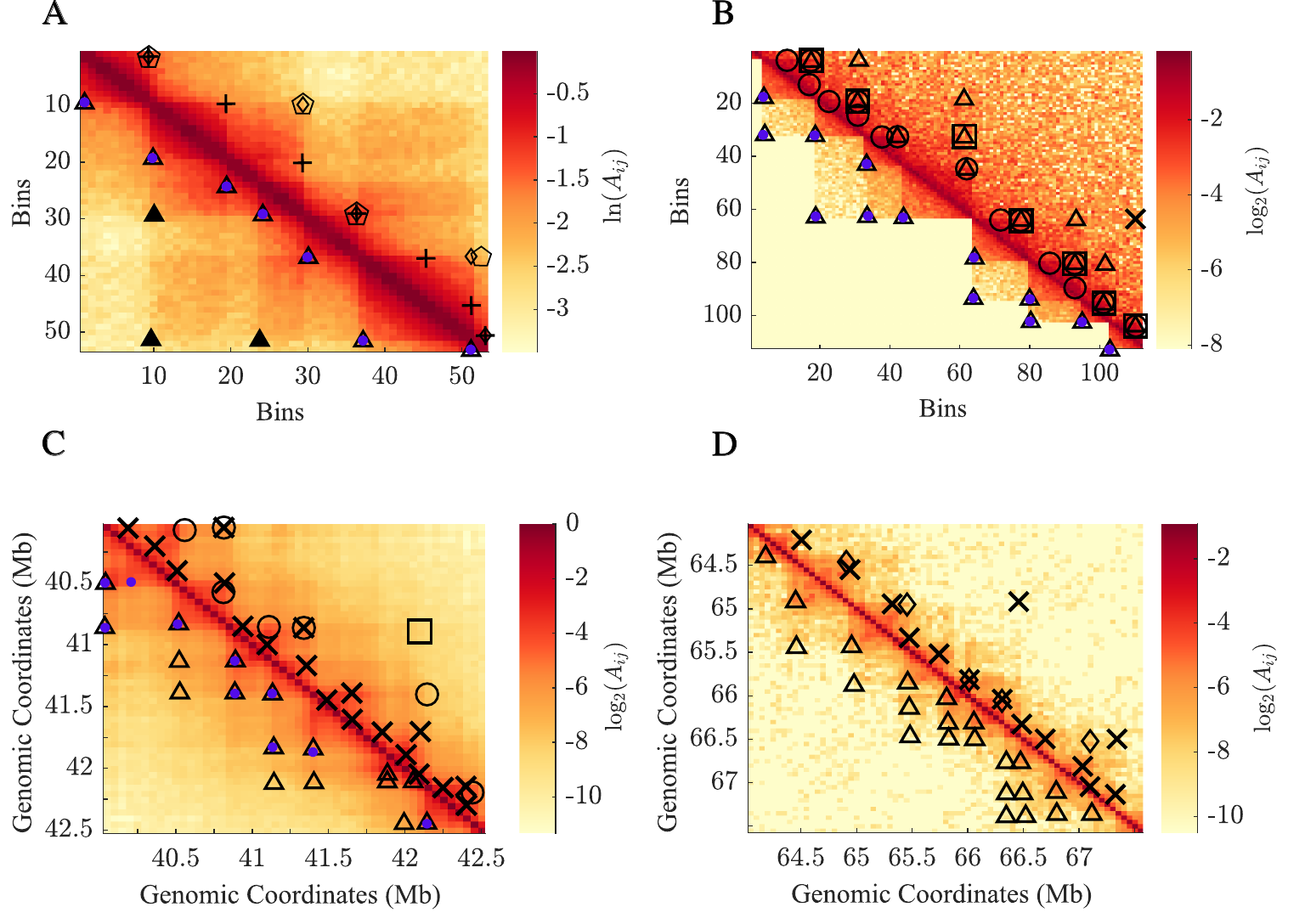}
\caption{{\bf Demonstration of TAD predictions across four different types of Hi-C maps.}
Simple synthetic Hi-C map (on $\ln$ scale) with TAD predictions from the four tested TAD identification algorithms. We show the predicted TADs for KerTAD (open triangles) and ground truth (blue circle) in the lower triangular matrix. The predicted TADs for HiCSeg (open pentagons), TopDom (open diamonds), and CHDF (crosses) are shown in the upper triangular matrix. Gray triangles are examples of TADs that KerTAD identifies if the restriction of one TAD per row for simple maps is removed. B: Complex synthetic Hi-C map (on $\log_2$ scale) with added noise, $\chi=1$. The upper triangular matrix shows the TAD predictions for all four tested algorithms on the noisy Hi-C map: KerTAD (open triangles), deDoc (crosses), Armatus (open squares), and SpectralTAD (open circles). The lower triangular matrix shows the same Hi-C map with no added noise, $\chi=0$, ground truth (blue circles), and the predictions of KerTAD. C: Manually annotated GM12878 chromosome $22$ Hi-C map from genomic coordinate $40$ to $42.5$ Mb at 50kb resolution. Predictions from the same TAD identification algorithms in B are shown in the lower triangular matrix. D: Non-annotated human lymphoblastoid Hi-C map with predictions from the same TAD identification algorithms as C.}
\label{Fig9}
\end{figure}

\section*{Discussion}

In this article, we developed a novel algorithm, KerTAD, to identify TADs in Hi-C maps. Most previous TAD calling algorithms assume simple Hi-C maps, i.e. each diagonal element of ${\cal A}_{ij}$ must belong to one and only one TAD. For simple Hi-C maps, when a TAD is identified at element $i$ and $j$, the next TAD must have a starting index of $j+1$ and there can be no additional TADs between $i$ and $j$.  In contrast, our method does not assume that Hi-C maps are simple and can accurately identify nested, overlapping, and gapped TADs. Among the few algorithms that can identify TADs in complex Hi-C maps, which is necessary for accurate TAD identification in experimental Hi-C maps, there is a large discrepancy in the number and size of TADs called, even among replicate Hi-C maps from the same experiment. Here, we present a novel algorithm that consistently outperforms other TAD identification algorithms on synthetic and manually annotated Hi-C maps, while being robust to noise and sparsity. 

KerTAD uses two kernel-based techniques that detect complementary features of Hi-C maps. The method focuses on regions of Hi-C maps near the diagonal where there are large changes in intensity and strong corner points. We show that KerTAD outperforms six state-of-the-art TAD identification algorithms on both synthetic and manually annotated experimental Hi-C maps. In particular, we calculate the TPR and FDR by comparing the results for the predicted TADs for each algorithm to ground truth for the synthetic and experimental manually annotated Hi-C maps. We also test the performance of the TAD identification algorithms on complex, synthetic Hi-C maps with increasing levels of impulse noise and sparsity. For all of the Hi-C maps with ground truth that we tested (i.e. simple and complex synthetic, noisy and sparse, and manually annotated, experimental), our method has the highest TPR and negligible FDR. 

We also find that our method has low variance in the median number and size of TADs across replicates for the experimental Hi-C maps without ground truth. In previous work ~\cite{bib27,bib28} that evaluated TAD identification algorithms, algorithms that can identify nested and overlapping TADs predict more TADs and possess higher variance in the number of identified TADs over replicates. This result is consistent with the fact that simple TAD identification algorithms can only call at most $N$ TADs for a Hi-C map with $N\times N$ elements, whereas algorithms for complex Hi-C maps can identify at most $N^2$ TADs. Our results also show that algorithms for complex Hi-C maps identify more TADs than those for simple Hi-C maps, e.g. deDoc identifies significantly more TADs and with higher variance among replicates than TopDom. However, unlike deDoc, our method, which can identify TADs in complex Hi-C maps, shows significantly lower variation among replicates, with maximum and minimum values for the numbers and sizes of TADs comparable to those for TopDom. The fact that our method generates results for the numbers and sizes of TADs with small variations among replicates suggests that our method identifies the most important features of Hi-C maps that are insensitive to resolution and downsampling.

While KerTAD outperforms other current TAD identification algorithms on synthetic Hi-C maps, it can be improved. For Hi-C maps where there are high-intensity regions compared to the local neighborhood, we find that despite TVR reducing the variation, our method still tends to identify TADs in the regions of high intensity, rather than in regions of low intensity. Since TADs are usually defined \emph{locally}, using global techniques that threshold across the whole Hi-C map will invariably suffer from this problem. Unfortunately, this results in a well-known dilemma: if one does not normalize weaker intensity regions, the algorithm will miss TADs, but normalizing weak intensity regions will bring out noise causing false positive TADs. This can be controlled to some degree by separating large maps into smaller ones (setting $\gamma = 1$) but risks "cutting off" TAD boundaries. In future work, we will develop new techniques to reduce noise, while maintaining the ability to identify TADs in weak intensity regions.

Because our method possesses the highest accuracy on synthetic and manually annotated experimental Hi-C maps, we hypothesize that our method will be accurate in capturing the true number and size of TADs in experimental Hi-C maps. However, it is worth reiterating that there is currently no ground truth definition of TADs in experimental Hi-C maps, which means that TPR and FDR on synthetic and manually annotated data, while useful, are only proxies for the accuracy of TAD identification algorithms on experimental Hi-C maps. Previous research groups ~\cite{bib26,bib27,bib28,bib30} have benchmarked their TAD identification algorithms using different metrics. For example, several studies have searched for correlations between predicted TAD boundaries and CTCF enrichment as a measure of TAD identification accuracy. However, this benchmark may not be related to benchmarks that rely on visual identification of TADs in experimental Hi-C maps. 

Currently, there can be large variations in the experimentally determined Hi-C maps from one experiment to the next.  
As chromatin conformation capture experiments continue to improve, it will be possible to determine well-defined, relatively noise-free, and experimentally reproducible Hi-C maps. It is also important to understand how Hi-C maps depend on the phase of the cell cycle, cell type, cell-to-cell fluctuations, and tissue type in each organism. After such experimental studies are carried out and well-defined Hi-C maps are obtained, computational studies can be carried out to determine in an unsupervised way the important features that distinguish one Hi-C map from another.  After identifying these key features, further studies can be carried out to understand the spatiotemporal dynamics of chromatin that give rise to each of the key features in Hi-C maps. 
 
\section*{Acknowledgments}
The authors acknowledge support from NSF Grant No. 1830904 (L.M., M.C.K., S.G.J.M., and C.S.O.). This work was also supported by the High Performance Computing Facilities operated by Yale's Center for Research Computing. 
\\
\vspace{16mm}

\section*{Supporting information}
\begin{figure}[!h]
\includegraphics[width=\textwidth]{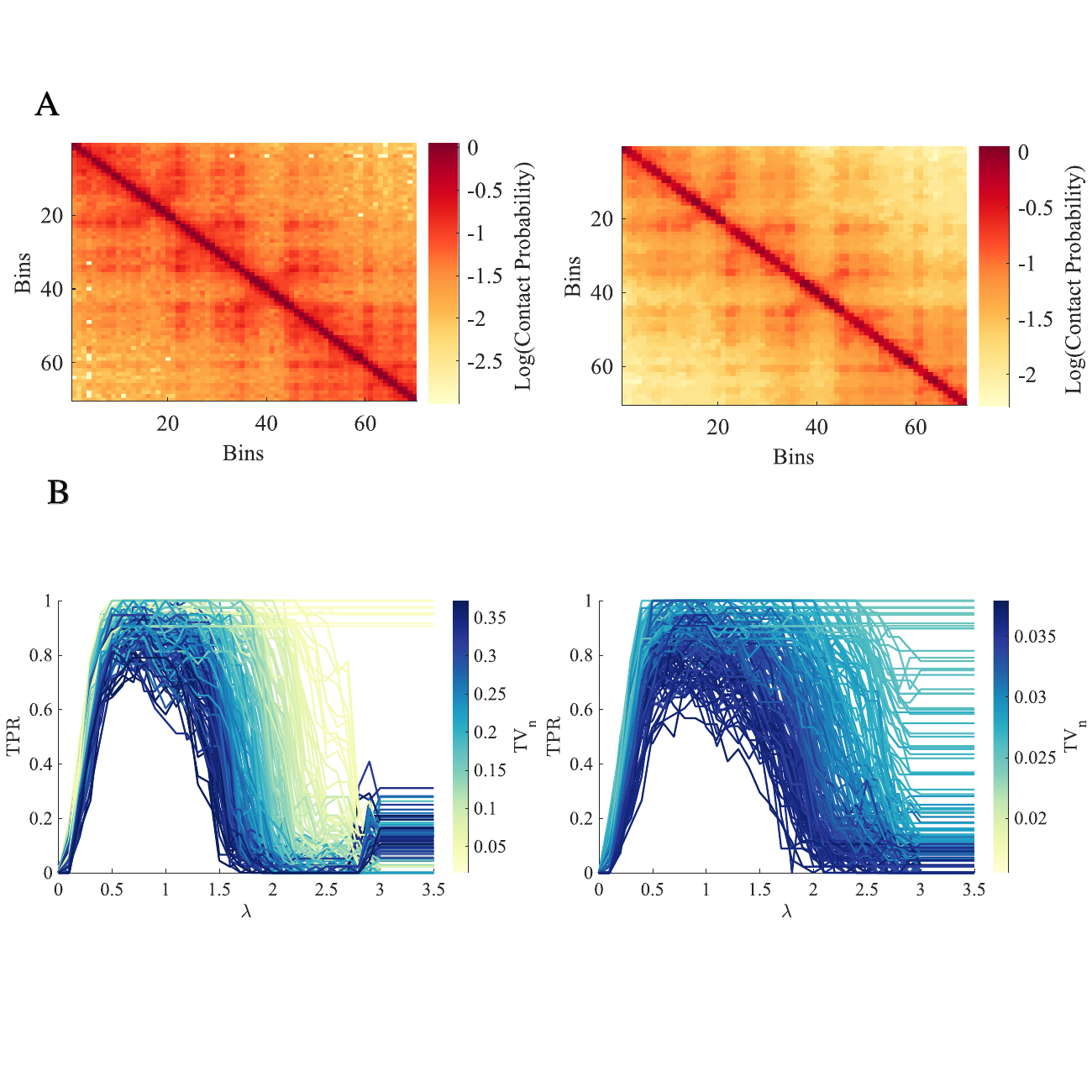}
\caption{{\bf Regularization parameters} A: Hi-C map of meiotic S. cerevisiae (left) before and (right) after total variation regularization with $\lambda =1$. (See Eq.~{\ref{eq:constraint}.)}
B: (left) $\lambda$ plotted versus the true positive rate (TPR) for 210 synthetic Hi-C maps generated by the block copolymer molecular dynamics simulations, with each map having a varying amount of impulse noise added, which increases the total variation. Each line represents one Hi-C map and the color represents the normalized starting total variation ($V_N= V/N^2$, where $V$ is given by Eq.~\ref{eq:tv} and $N$ is the size of $\mathcal{A}_{ij}$), before total variation regularization is applied. We find a peak in the accuracy (TPR) near $\lambda =1$ and no change in TPR for $\lambda>3$. (right) Similar plot to that shown on the left, but the sparsity $\xi$ is tuned to increase the total variation. }
\label{FigS1}
\end{figure}

\end{document}